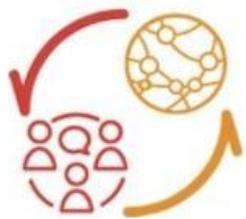

https://indcor.eu

# A shared vocabulary for IDN

**INDCOR white paper 1**

Version 1.0

Hartmut Koenitz[1], Mirjam Palosaari Eladhari[2], Sandy Louchart[3], Frank Nack[1]

[1] University of Amsterdam, The Netherlands
h.a.koenitz@uva.nl; nack@uva.nl
[2] Stockholm University, Sweden
mirjam@dsv.su.se
[3] Glasgow School of Art, UK
s.louchart@gsa.ac.uk

### Authors of sample entry
Christian Roth[1,] Elisa Mekler[1],

[1] HKU University of the Arts Utrecht
christian.roth@hku.nl
[2] Aalto University, Finland
elisa.mekler@aalto.fi

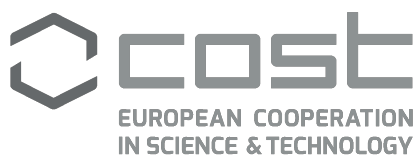

# TABLE OF CONTENTS

# Executive Summary

COST Action 18230 INDCOR (Interactive Narrative Design for Complexity Representations) is an interdisciplinary network of researchers and practitioners intended to further the use of interactive digital narratives (IDN[1]) to represent highly complex topics. IDN possess crucial advantages in this regard, but more knowledge is needed to realize these advantages in broad usage by media producers and the general public. The lack of a shared vocabulary is a crucial obstacle on the path to a generalized, accessible body of IDN knowledge. This white paper frames the situation from the perspective of INDCOR and describes the creation of an online encyclopedia as a means to overcome this issue. Two similar and successful projects (The Living Handbook of Narratology and the Stanford Encyclopedia of Philosophy) serve as examples for this effort, showing how community-authored encyclopedias can provide high-quality content.

The authors introduce a taxonomy based on an overarching analytical framework (SPP model) as the foundational element of the encyclopedia, and detail editorial procedures for the project, including a peer-review process, designed to assure high academic quality and relevance of encyclopedia entries. Also, a sample entry provides guidance for authors.

The whitepaper is intended as a foundation for future work in INDCOR and as orientation for the field of interactive narrative research and is openly available. In particular, it will serve as orientation for contributors to the IDN encyclopedia and should be widely shared.

---

[1] IDN should be read in the plural form (Interactive Digital Narratives), except when it is a modifying adjective, in which case it should be read as singular (Interactive Digital Narrative)

# 1    INDCOR – Goals and Approach

The aim of the multinational research network INDCOR is to facilitate the use of interactive digital narratives (IDN) to represent highly complex topics, e.g. global warming, the refugee situation in the EU, pandemic events such as COVID19, or the switch to E-mobility. By means of more complex representations, IDNs can accommodate competing perspectives, as well as choices and diverging outcomes, thus addressing the issue of fake news and the growing complexity of e-argumentation in social media.

> Interactive Digital Narratives (IDN) are "expressive narrative forms implemented as a computational system [...] and experienced through a participatory process" (Koenitz 2015) which can be realized in many different ways, including video games, interactive documentaries and mixed reality projects.

> IDN have specific characteristics that make them an ideal platform for the representation of complex topics:
> - IDN create active audiences by providing agency – the ability to influence the continuation and outcome of an experience – and thus improve understanding through self-directed experiences;
> - IDN can contain many different – even competing – perspectives in the same artefact. They are not bound by the practical size limitations of traditional media;
> - IDN can encode complex dependencies and allow them to be experienced as narratives – the fundamental human form of communication.

So far, the potential of IDN for representing complex topics has been largely untapped, due to a number of factors - chiefly amongst them the fragmentation in both IDN research and practice, issues which INDCOR addresses.

> INDCOR is a coordinated, interdisciplinary research effort to lay the foundations for an IDN research field in general and more specifically to establish IDN best practices in design, production, and application for complexity narratives.

Towards this goal, the INDCOR network will produce insights applicable to the design process, production/distribution aspects, audience reaction, societal context and educational uses. Concretely, the INDCOR approach includes analyzing the many notable IDN works in existence, empirical evaluations of successful design approaches, audience research, and analyses of the societal context, but also workflow analysis to understand the best integration of IDN into existing journalistic practices. While some of these topics have been investigated before, due to fragmentation and even more so because of varying terminology, they did not result in a generalized body of knowledge about IDN. INDCOR's first goal is therefore to establish a shared vocabulary.

> A shared vocabulary is a fundamental condition for creating shared knowledge. INDCOR is establishing one for IDN research and practice.

INDCOR is not the first project to take an in-depth look at IDN. The IRIS (Cavazza et al. 2008) and RIDERS (Aylett, Louchart, and Weallans 2011) projects also approached IDN from a networking perspective. Both projects aimed to identify and determine principles and processes towards an understanding of multidisciplinary research in the context of IDN. RIDERS focused on exploring creative interactivity while IRIS provided a more formal description of IDN systems and approaches along with a glossary of terms. INDCOR builds on these foundations through its membership and acknowledgement of prior work conducted in the domain. INDCOR is however focused on the representation of complexity and consequently encompasses a more diverse variety of disciplines than these prior efforts. As such, a shared analytical vocabulary providing a contextualized representation of the IDN landscape is a crucial factor towards allowing different disciplines to meaningfully contribute to the development of the IDN domain and to the representation of complex issues and topics.

## 2    INDCOR - Structure

Having started in late 2019, the INDCOR project divided work into four work groups, WG1 Design and Development, WG2 Conceptualizing Narrative Complexity, WG3 Evaluation, and WG 4 Societal Context.

**WG1 Design and Development** is concerned with aspects of making - the design of the artifact and its development in an authoring tool, or by working with a programming language directly. The central focus is the design of IDN prototypes as well as the analysis of the design and evaluation process with the aim to make the resulting knowledge available to practitioners for the representation of complex topics. A first target group will be journalists and media organizations. Further targets can be broad and will include artistic and entertainment domains.

**WG2 Conceptualizing Narrative Complexity** is concerned with a conceptual framework to understand interactive digital narratives. This workgroup also leads the effort for the creation of a shared vocabulary. WG2 will investigate, develop and disseminate concepts intended to better understand the elements and context of IDN, especially in relation to complex topics. The output of WG2 provides a foundation for IDN research (e.g. on specific subject matters and the ways in which participants are involved in a complex, communicative interaction), IDN applications (e.g. in journalism or education), and IDN training in higher education and in the private sector.

**WG3 Evaluation** collects and applies evaluation methods for understanding IDN. WG3 aims to gain a holistic and transdisciplinary understanding of the moment-to-moment, short term, and long-term outcomes of engagement with interactive narratives for complexity representation. WG3 will investigate the impact of IDN complexity representations in audiences, in a manner that productively informs the other work groups and different stakeholders. Specifically, we identify and establish key success indicators, collect and develop methods and frameworks to assess and evaluate interactive narratives, as well as analyze the impact of different aspects and manifestations of interactive narratives representing complex topics.

**WG4 Societal Context** considers the impact of IDN complexity representations in society. WG4 focusses on societal contexts in order to better understand the connections between contexts, IDN, and complexity. In more abstract terms, WG4 examines the relationship between IDN and their external environments. Thereby, WG4 aims to refrain from privileging particular contexts over others and to transcend traditional binary divides that are made, for instance, between micro and macro levels of narratives and between their discursive and material elements. At the same time, WG4 aims to attend to the intersections between different contexts in order to prevent myopic viewpoints or polarization.

These different foci necessarily mean the applications of different methods and approaches, and this brings the danger of misunderstandings. This is true on multiple levels, both within workgroups (due to different disciplinary origins of group members) and between workgroups (due to the different objects of study). The lack of a vocabulary shared across the different work groups is therefore a key issue that the network intends to solve. What follows is the network's plan in addressing the issue.

## 3   Addressing the Lack of a Shared Vocabulary for IDN

> INDCOR will address the lack of a shared vocabulary through the creation of a living encyclopedia of IDN vocabulary

The lack of a shared vocabulary is a longstanding issue of the field of interactive digital narrative (Koenitz, Ferri, and Sezen 2009; Koenitz et al. 2013; Koenitz 2014; Koenitz 2016; Thue and Carstensdottir 2018). The root of this problem is the fact that scholars and practitioners concerned with the topic of interactive narrative originate in a number of different fields, including literature studies, film studies, computer sciences (both from an Artificial Intelligence (AI) and an Human Computer Interaction (HCI) perspective), media studies, creative practices and many more. All of these fields have associated specific vocabulary, positioned within a semantic field of meaning developed in their respective tradition and thus often not immediately accessible to "outsiders" from a different field. The issue is further aggravated by the fact that many common terms used in Interactive Digital Narrative research and practice – such as "narrative", "plot" or "story" – have both a common meaning in everyday conversation, and also specific ones in scholarly and professional contexts. The 'story' of a journalist is not exactly the same as the 'story' of a film director and what is exactly meant in each case is only fully accessible to practitioners in the respective fields. Equally, 'narrative' in the sense used by sociology scholars describing a "group narrative" is not the same ontological entity that an AI researcher concerned with procedural generation has in mind. In a recent article Koenitz and Eladhari compared this status to the biblical metaphor of the "Babylonian confusion" (2019).

This issue of diverse-language-use manifests as a considerable obstacle to productive work, especially in interdisciplinary settings, for example in the multinational research network INDCOR. Here, a shared vocabulary is a foundational necessity in order to enable scholars and practitioners from various disciplines to meaningfully contribute to IDN research, further the development in professional settings (design practices, production workflows), spur wider adoption and thus realize market opportunities. In addition, the question of a shared vocabulary is also a crucial element in the networks' intention to improve the understanding of IDN in society at large. The question is thus - both for the research network INDCOR and for the overall field of IDN research: how to overcome the Babylonian confusion?

In this paper, we propose to address the issue through the creation of a "living encyclopedia of IDN vocabulary" based on an overarching analytical framework (SPP model) and associated taxonomy. We detail the process for the creation and continued development of such a resource within INDCOR and also invite the community to participate in the development of this central aspect for the fledgling field of IDN research and practice.

## 4    Foundational Considerations for a Shared Vocabulary

Before getting into the details of our proposal, we would like to describe our process and consider the aim and purpose of a shared vocabulary to determine its scope.

> A shared vocabulary means that key terminology is understood across a multidisciplinary field.

In order to enable such a shared understanding, an overarching analytical perspective is necessary. This is an insight gained by our own experience in the INDCOR project. The setup of workgroups was a bottom-up approach developed by the expert community at the first meeting in Brussels. As work progressed, a long list of terms to be defined in a shared vocabulary was quickly produced, but subsequent discussions made clear that a consensus on how to connect different definitions was elusive. Consequently, we identified the need of an overarching abstraction to guide our work on a shared vocabulary of IDN. Without it, there was a manifest risk that a shared vocabulary would replicate the existing Babylonian confusion and thus miss its central aim. In ongoing work and subsequent meetings in Vienna and online, the SPP (System Process Product) model (Koenitz 2015) was the one singled out across workgroups that could function as a starting point for connecting the four pillars that are expressed as workgroups in INDCOR.

> There is a need for an overarching abstraction. Koenitz' SPP model (System Process Product) serves in this role for INDCOR.

### 4.1 An Overarching Analytical Perspective: the SPP Model

The SPP model is a media-specific perspective that identifies three broad categories for the analysis of IDN artefacts, reflecting its different stages: *system* – the digital artifact, *process* – the interactive experience of a system, and *product* – the result of the experience, either in the form of a recording or as a retelling (Eladhari 2018) to others (figure 1).

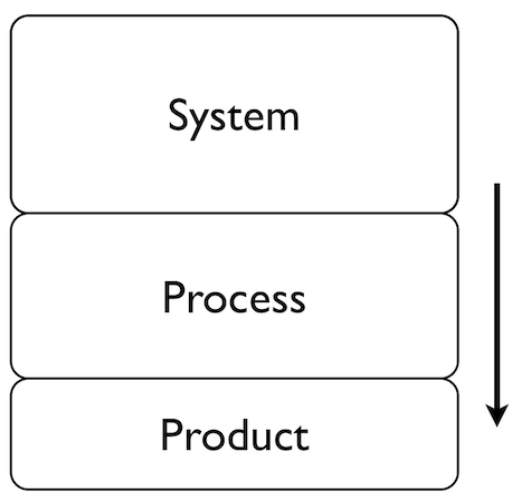

Figure 1. SPP model

The SPP model takes the systemic, dynamic character of IDN works as its central characteristic, building on a foundation laid by cybernetics (Wiener 1948), and cybernetic art theory (Ascott 1964; Ascott 1968), as well as earlier perspectives on interactive forms of narration (Laurel 1986; Jennings 1996; Murray 1997; Montfort 2005; Murray 2011). In order in order to understand the specific aspects of IDN and avoid limitations inherent in adapted perspectives[2], the SPP model does not rely on underlying models derived from the classical formal study of literature and the cinema in narratology. Instead, it acknowledges the 'cognitive turn' in narratology – a perspective that understands narrative not as a property of certain types of artefacts, but as a cognitive function, a "frame for constructing, communicating, and reconstructing mentally projected worlds" (Herman 2002). This perspective opens up a space for novel kinds of narrative manifestations – as in principle any artefact can be considered a narrative as long as it triggers the cognitive frame of narrative. In other words: an IDN work does not need to be similar to the literary novel or the movie to be considered a narrative. This perspective creates opportunities for both theoretical development and novel practices.

> An IDN work does not need to be similar to the literary novel or the movie to be considered a narrative

Thus, the SPP model is concerned with defining aspects of IDN *systems*, their *processes* and resulting *products* and organizes related concepts and design aspects accordingly. For instance, *system* contains the *protostory*, the sum of all potential narratives that can be instantiated with a given artefact (figure 2). Further aspects related to an IDN work (*narrative design, user interface, assets, environment definitions/rule systems*) are subcategories of *protostory*. Conversely, concepts, practices and examples related to the manner in which an IDN is presented to an audience (i.e. visual presentation, interaction, feedback) would be represented within the category of *process*. This would mean for example that an AI engine is described as a *rule system* within the protostory and its particular implementation as a part of the *narrative design*. Finally, *product* describes the output of a *process*, either as a recording (objective product) or as re-telling (Eladhari 2018) (subjective product).

---

[2] Cf. N. Katherine Hayles' call for a "media-specific analysis" of digital forms of narration (Hayles 2002) and Hausken's warning of "media blindness" (2004).

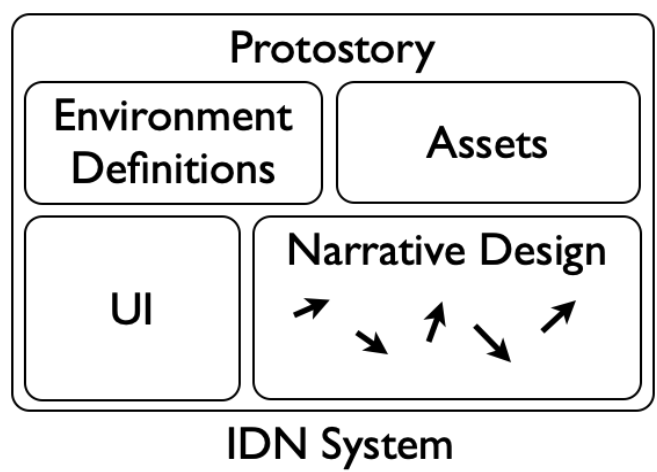

Figure 2. Protostory and its elements

For the INDCOR shared vocabulary project, the SPP model is used as a foundational analytical framework to connect central concepts of IDN in terms of design and development, theory building, societal context and evaluation. This choice is pragmatic, based on several advantages:

- SPP emphasizes the specificity of this dynamic form thus providing a clear distinction from earlier, fixed forms of narrative such as the printed novel or film and associated vocabulary and thus avoids the pitfalls of re-defining existing terminology;
- The SPP model takes the systemic nature of IDN artifacts as foundational, building on a solid lineage of cybernetics, cybernetic art and system theory;
- The SPP model continues efforts by scholars such as Pamela Jennings, Brenda Laurel, Janet Murray and Nick Montfort in understanding the specific aspects of Interactive Digital Narratives;
- The SPP model features an inclusive view that acknowledges the wide variety of different forms of IDN, including hypertexts, journalistic interactives, narrative-focused video games, interactive documentaries, installation pieces, and AR/VR work as well as emerging forms;
- The SPP model provides a high-level model of IDN works and their relationship with their audiences that is open both to extensions and

further lower-level specification and thus the SPP model can serve as a central element in an IDN taxonomy.

> As a foundational abstraction, the SPP model emphasizes specificity, avoids the pitfalls of adapted terminology and enables precise definitions of the phenomena at hand

In the next section, we describe the SPP model in more detail, before outlining our specific approach on implementing a shared vocabulary.

The conceptual framing provided by such a model is particularly important when considering the crucial role we expect disciplines adjacent to IDN to play in the development and establishment of the shared vocabulary. Whilst a theatre writer and a game developer might differ in their definitions of a story environment, their definitions could be presented alongside each other in a sub-section of this particular model. Conversely, a film director's definition of an environment, might be more oriented towards the process or product aspect of SPP and closer related to the audience experience than the story setting commonly observed in games and theatre.

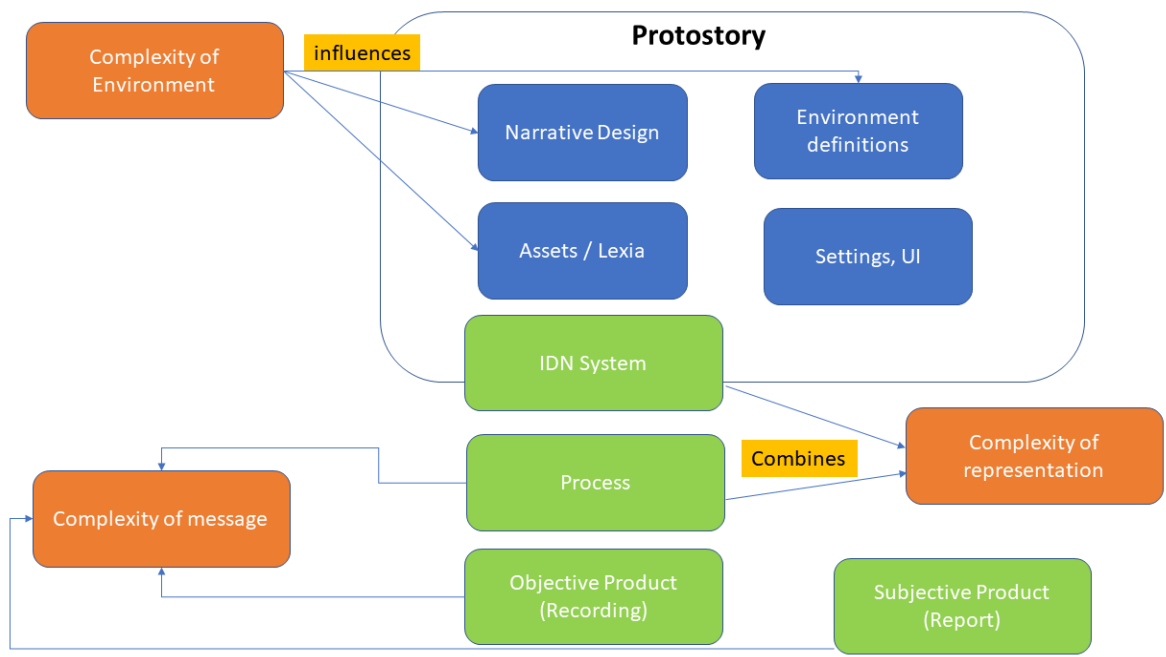

Figure 3 - A contextual framework for a shared vocabulary

The diagram in Figure 3 provides a framework through which IDN and complexity can be contextualized using the generic IDN elements identified by the SPP model.

## 4.2 A Starting Point for the Field

Any choice of overarching analytical perspective is open to criticism and will be controversial to some researchers and/or professionals, yet there is the simple fact that a starting point must be chosen and given the above-mentioned advantages, the SPP model provides a solid foundation.

> A shared vocabulary will not make all difference disappear, but it can serve as a coordinate system for a field

A shared vocabulary does not mean that all differences in meaning or historical context would simply disappear, or that scholarly dispute all of a sudden ends, but that a coordinate system would be established against which extensions and alternative views can be discussed and understood. Conversely, from this clear vantage point, explicit connections can be drawn which will enable scholars to better understand each other and productively work together. Most importantly, since we aim to make the concept of a shared vocabulary available for the whole community of researchers and designers, it can serve as a hub for knowledge exchange and an important step in building a field accessible also to newcomers and related disciplines.

## 5   A Taxonomy for IDN

The SPP model is concerned with analyzing the IDN artifact, which means it does neither explicitly cover the conditions leading to a work (Ideation) nor its creation process (Authoring). Conversely, it also does not concern itself with societal effects of the work in question or on other works (Critical discourse). As a broad framework, it also does not provide the granularity necessary for an encyclopedia. Consequently, the authors of this whitepaper have developed a taxonomy for interactive digital narrative on the basis of the SPP model with the top-level categories of *authoring*, *artefact* and *critical discourse* (table 1).

> The authors have developed a taxonomy on the basis of the SPP model to serve as the foundation for the IDN encyclopedia

The taxonomy relates authors' contributions to an overall structure. For example, *transformation* is categorized as an aesthetic quality and thus an aim during the authoring process (**Authoring** > Ideation > Content > Aesthetic qualities >Transformation) and an element of the experience of successful design (**Artefact** > Process > Experience > Aesthetic > Transformation)

We understand this taxonomy as a first effort (version 1.0) explicitly open to changes and amendments as the result of discussions and developments in INDCOR as well as in the research and practice communities at large.

> The taxonomy is open to changes and amendments - it therefore has a version number attached (currently 1.0)

The taxonomy itself represents the core of the shared vocabulary and as such provides a foundation which supports the community-driven effort in developing a more extensive collection of terminology. This explicitly means that further terms should be proposed and will be integrated to grow the vocabulary. The focus here will be on concepts not already well-defined in existing collections, e.g. *The Living Handbook of Narratology* (described below in section 6.1).

This taxonomy also provides a starting point for an IDN Ontology understood in the sense used in computer sciences, as a structure that defines its members not only through self-contained definitions, but also through connections as part of a network.

## 5.1 IDN Taxonomy V 1.0

1. **Authoring**
    1.1. Ideation
        1.1.1. Affordances
            1.1.1.1. Procedural
            1.1.1.2. Participatory
            1.1.1.3. Spatial
            1.1.1.4. Encyclopedic
        1.1.2. Audience
            1.1.2.1. Social
            1.1.2.2. Private
            1.1.2.3. Expectations
        1.1.3. Content
            1.1.3.1. Complexity
                1.1.3.1.1. Topic
                1.1.3.1.2. Addressee
                    1.1.3.1.2.1. Social
                    1.1.3.1.2.2. Private
            1.1.3.2. Prior Narratives
            1.1.3.3. Material
                1.1.3.3.1. Fiction
                1.1.3.3.2. Non-fiction
            1.1.3.4. Form
                1.1.3.4.1. Interactive Documentary
                1.1.3.4.2. Video game
                1.1.3.4.3. Hypertext fiction
                1.1.3.4.4. Location-based
                1.1.3.4.5. AR/VR
                1.1.3.4.6. Mixed
            1.1.3.5. Aesthetic qualities
                1.1.3.5.1. Immersion
                1.1.3.5.2. Agency

- 1.1.3.5.3. Transformation
- 1.1.3.6. Meaning Making
  - 1.1.3.6.1. Mental processes
    - 1.1.3.6.1.1. Hermeneutic circle
    - 1.1.3.6.1.2. Narrative cognition
    - 1.1.3.6.1.3. Cognitive reduction
    - 1.1.3.6.1.4. Embodied cognition
  - 1.1.3.6.2. Rhetoric
  - 1.1.3.6.3. Interface
    - 1.1.3.6.3.1. Interaction Metaphor
  - 1.1.3.6.4. Prediction of Audience reaction
    - 1.1.3.6.4.1. Feedback
- 1.2. System Implementation
  - 1.2.1. Protostory
    - 1.2.1.1. Asset creation
      - 1.2.1.1.1. Characters
      - 1.2.1.1.2. Props
    - 1.2.1.2. Environment building
      - 1.2.1.2.1. Geographic
        - 1.2.1.2.1.1. Landscapes
        - 1.2.1.2.1.2. Buildings
      - 1.2.1.2.2. Rule Systems
        - 1.2.1.2.2.1. Physics Systems
        - 1.2.1.2.2.2. Societal Rules
    - 1.2.1.3. UI/Interface building
    - 1.2.1.4. Interactive Narrative Designing
      - 1.2.1.4.1. Combinatorics
      - 1.2.1.4.2. Structure
        - 1.2.1.4.2.1. Events
        - 1.2.1.4.2.2. Narrative Vectors
      - 1.2.1.4.3. Experience Schema
        - 1.2.1.4.3.1. Narrative Cognition
        - 1.2.1.4.3.2. Narrative Experience

- 2.1.1.3.2.1. Physics Systems
- 2.1.1.3.2.2. Societal Rules
- 2.1.1.3.3. UI/Interface
- 2.2. Process
    - 2.2.1. Participation
        - 2.2.1.1. Interaction
            - 2.2.1.1.1. Active/Performance
            - 2.2.1.1.2. Passive/Sensoric
        - 2.2.1.2. Sense Making
            - 2.2.1.2.1. Double Hermeneutic circle
                - 2.2.1.2.1.1. Reflection
                - 2.2.1.2.1.2. Action
                    - 2.2.1.2.1.2.1. Planning
                    - 2.2.1.2.1.2.2. Execution
        - 2.2.1.3. Experience
            - 2.2.1.3.1. Aesthetic
                - 2.2.1.3.1.1.1. Immersion
                - 2.2.1.3.1.1.2. Agency
                - 2.2.1.3.1.1.3. Transformation
            - 2.2.1.3.2. Rhetoric
- 2.3. Product
    - 2.3.1. Objective (Recoding)
        - 2.3.1.1. Interaction analysis
        - 2.3.1.2. Attention analysis
        - 2.3.1.3. Pace control
    - 2.3.2. Subjective (Retelling)
        - 2.3.2.1. Experience Model
            - 2.3.2.1.1. Narrative cognition
            - 2.3.2.1.2. Cognitive reduction
            - 2.3.2.1.3. Embodied cognition
        - 2.3.2.2. Structure Inference
        - 2.3.2.3. Updating episodic memory
        - 2.3.2.4. Updating perception memory

3. **Critical Discourse**
   3.1. Effect
      3.1.1. Comparison Intent/Experience
   3.2. Reflective Analysis
      3.2.1. Methods

## 5.2 Example Definition

For an example entry see Appendix 1 on Transformation

# 6   A (Living) Encyclopedia for IDN

With the foundations of analytical foundation and taxonomy in place, the question is how to make it accessible and enable community involvement and further development. Examples for accessible shared vocabularies exist in the form of online encyclopedias and this is the model we have decided to follow. Arguably the most successful example for general knowledge is Wikipedia (2020b), the free online encyclopedia. In the scholarly realm, two particularly successful examples are the Living Handbook of Narratology (Hühn 2015) and the Stanford Encyclopedia of Philosophy (2020a). Both Wikipedia and the scholarly resources are viable models. We aim to follow the latter model for a variety of reasons and will be developing an online encyclopedia for IDN research and practice.

An argument for creating an encyclopedia rather than a shared Wikipedia is that, based on the examples above, these endeavors have proven more successful in terms of creating a complete and high-quality curated result. The reasons for this outcome are likely multifold, but we speculate that an encyclopedia entry is more rewarding for authors to participate in, since their efforts are clearly recognizable and properly credited. As for its audience, an encyclopedia appears to be more curated in both its individual content, and as a whole, since an encyclopedia by its nature promises to provide a more comprehensive view of a given topic, rather than to rely on seemingly random selections and emphasis determined by wiki-authors' preferences. This being said, the two methods are not mutually exclusive. A Wikipedia site could likely serve well as a pre-stage in the production of the Encyclopedia of IDN the way we envision it.

## 6.1 The Living Handbook of Narratology

The Living Handbook of Narratology (LHN) can be read as a book. It consists of chapters and sections which explain in depth a set of concepts that are chosen by the curating editorial board. LHN originated as a printed book in 2009. Content was added in a Wikipedia format up until 2013, after which it changed to the current form of an encyclopedia. At the time of writing (June 2020) LHN consists of 68 concepts, each making up a longer section.

The article text about each concept is most often attributed to one author, sometimes to two or three. For example, the text on the concept Reader (see Figure 1) is authored by renowned narratology scholar Gerald Price, and follows the recipe of what an entry in the LHN should consist of: Definition, Explication, a number of content specific sections, bibliography, works cited, and further readings.

Figure 1: Article in the Living Handbook of Narratology

The articles in LHN are of substantial length, often exceeding 4000 words. The 68 concepts included in the LHN are chosen as being central specifically for the field of narratology as it is understood by the editorial board.  If a similar approach was adopted for creating an encyclopedia for IDN, a likely - corresponding - article of a 'reader' would be that of the concept of the Interactor, eg the person or implied person who traverses an IDN. The original work on the LHN was supported by a grant from the German Research Association (DFG).

## 6.2 The Stanford Encyclopedia of Philosophy

The Stanford Encyclopedia of Philosophy (SEP) (2020a) is an established source in the field of Philosophy, featuring 1600+ entries. It was designed from the onset to be a high-quality resource which makes use of the specific advantages of an online publication with regular updates and revisions according to its own description:

> From its inception, the SEP was designed so that each entry is maintained and kept up-to-date by an expert or group of experts in the field. All entries and substantive updates are refereed by the members of a distinguished [Editorial Board](#) before they are made public. Consequently, our dynamic reference work maintains academic standards while evolving and adapting in response to new research. (2020a)

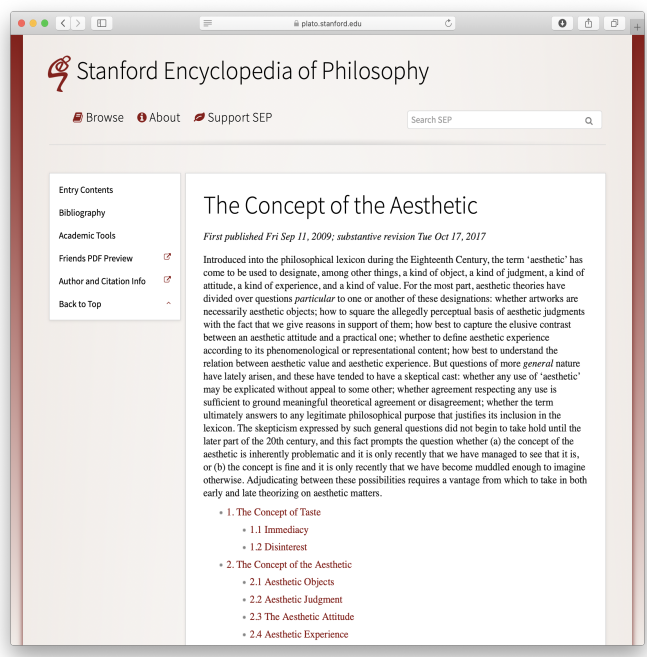

Figure 2: Article in the Stanford Encyclopedia of Philosophy

Articles in the SEP are prestigious and are of considerable length - often exceeding 10.000 words. Most entries are invited by the editorial board, although proposals are considered. The SEP maintains a quarterly publication/update schedule and entries might be cited including a version designation (e.g. "fall 2015). The SEP has been supported by seven grants over the course of nine years (2020a), mostly from the NEH (National Endowment for the Humanities), but also NSF (National Science Foundation) and the Andrew W. Mellon Foundation.

## 6.3 Editorial Procedures

Taking inspiration from the two examples, we see a strong editorial board as central to the success of this undertaking, providing guidance and assuring academic excellence. Conversely, we consider community engagement as crucial in making this undertaking a success, which means to aim at a good balance of both aspects. Therefore, the IDN encyclopedia will be composed of articles written by authors from the community which have undergone a thorough peer-review process. Articles will be published based on their academic rigor and the relevance of their contribution to the field.

The group of INDCOR chairs (nine scholars originating in different disciplines and at different career stages[3]) will act as the initial editorial board, soliciting authors from inside the project and the community at large for encyclopedia entries. At the time of writing, in June 2010, the INDCOR action consists of more than 140 scholarly experts and professionals concerned with IDN complexity representations. The editorial board would act as the initial group of reviewers, but will be inviting additional experts from the field to assure a diversity of perspectives and a high level of quality content also for topics outside their core expertise. The aim is to make the encyclopedia a high-quality resource where each entry is recognized as a publication in its own right and thus writing for the encyclopedia is a rewarding undertaking for contributors.

---

[3] Hartmut Koenitz, Mirjam Eladhari, Frank Nack, Agnes Bakk, Jose Manuel Noguera, Andrew Perkis, Sandy Louchart, Elisa Mekler, Lissa Holloway-Attaway

## 6.4 Content Structures of the IDN Encyclopedia

The proposed content structure of the Encyclopedia of IDN will need to evolve in tandem with the INDCOR project and developments in the research community.

In general, entries should be guided by the principles of multi-entry and interlinked dependencies. The INDCOR project provides a microcosm representing different perspectives in the IDN community, manifested in its four workgroups. Therefore, a guiding principle for every author is that they are writing from the perspective of one of the four workgroups, but need to consider how the same entry would be understood by, and thus be useful for, members from any of the other three workgroups. For example, if an entry is implicitly written from a design and development (1) perspective its content needs to provide 'entries' - hooks - that enable connections to how narrative complexity can be conceptualized and theorized (2), to how it can be evaluated (3), and to the question in what societal contexts it might apply to (4). Additionally, the encyclopedia needs to be able to represent the interlinked dependencies in the development of the IDN field. For example, that a specific work of interactive fiction might not have been possible without a certain authoring system, while the authoring system might not have been possible unless there had been precursors to it both in terms of theory and implementation.

The components of information that we are using as a starting point for encyclopedia entries include three main types of that would be linked via their (specific) content:

**Concept**

Relation to taxonomy (either defining a term that exist in the taxonomy or relating to closest existing term(s))

Alternative names for concept (if available)

Definition

Explication

History

Custom sections for specific concept

Bibliography

**IDN (a specific work)**

Relation to taxonomy

Precursors

Underlying system

Audience-facing elements (including interaction, audience feedback, discourse, cinematography, lighting etc)

Media specificity and traditions

Reception

**Late-breaking work**

This type of entry is for ongoing research, especially targeting current MA and PhD research work, but also ongoing and as of yet incomplete research projects. It uses the same categories as fully-developed entries, but ise marked as "late-breaking".

# 7   Conclusion and Future Work

In this paper, we have identified the lack of a shared vocabulary as a central issue for the INDCOR project as well as for the field of Interactive Digital Narrative in general. We further identified the need for an overarching analytical perspective, selected the SPP model in this capacity and have proposed to address this issue through a communal expert-authored and peer-reviewed encyclopedia. Examining the successful examples of the Living Handbook of Narratology and the Stanford Encyclopedia of Philosophy, we propose a structure for such an effort and introduce a taxonomy as a starting point to gather definitions of concepts central to IDN and useful across perspectives for scholars and practitioners in the field, taking into account four lenses: design and development, conceptualization and theory, evaluation, and societal context. We invite the research community to partake in the endeavor of creating a multifaceted, living encyclopedia providing a shared vocabulary for IDN. Participants in the INDCOR project will contribute to the development of a shared vocabulary through discussions and taskforce commitments within working groups activities and academic submissions to dedicated calls. The taxonomy presented in this document represents a framework through which working group activities will be developed, contextualized and coordinated. The INDCOR core management group, along with individual working groups, will set up a program of targeted interventions and open calls for INDCOR participants and the wider academic community to participate. As already mentioned, academic excellence is a key outcome for the INDCOR project and all contributions to the shared IDN vocabulary will be peer-reviewed and published based on their academic rigor and the relevance of their contribution to the field.

# References

Ascott, Roy. 1964. "The Construction of Change." *Cambridge Opinion*. Cambridge.

Ascott, Roy. 1968. "The Cybernetic Stance: My Process and Purpose." *Leonardo* 1 (2): 105. doi:10.2307/1571947.

Aylett, Ruth, Sandy Louchart, and Allan Weallans. 2011. "Research in Interactive Drama Environments, Role-Play and Story-Telling." In *Interactive Storytelling*, 1st ed., 7069:1–12. Lecture Notes in Computer Science. Berlin, Heidelberg: Springer, Berlin, Heidelberg. doi:10.1007/978-3-642-25289-1_1.

Cavazza, Marc, Stéphane Donikian, Marc Christie, Ulrike Spierling, Nicolas Szilas, Peter Vorderer, Tilo Hartmann, et al. 2008. "The IRIS Network of Excellence: Integrating Research in Interactive Storytelling." In *Interactive Storytelling: First Joint International Conference on Interactive Digital Storytelling, ICIDS 2008 Erfurt, Germany, November 26-29, 2008, Proceedings*, 5334:14–19. Lecture Notes in Computer Science. Berlin, Heidelberg: Springer Science & Business Media. doi:10.1007/978-3-540-89454-4_3.

Eladhari, Mirjam Palosaari. 2018. "Re-Tellings: the Fourth Layer of Narrative as an Instrument for Critique." In *Interactive Storytelling: 11th International Conference for Interactive Digital Storytelling, ICIDS 2018*, edited by Rebecca Rouse, Hartmut Koenitz, and Mads Haahr, 11318:65–78. 11th International Conference on Interactive Digital Storytelling, ICIDS 2018, Dublin, Ireland, December 5–8, 2018, Proceedings. Cham: Springer Berlin Heidelberg. doi:10.1007/978-3-030-04028-4_5.

Hausken, Liv. 2004. Coda. In Narrative Across Media (pp. 391–403). U of Nebraska Press.

Hayles, N. Katherine. 2002. *Writing Machines*. Cambridge: MIT Press.

Herman, David. 2002. *Story Logic*. Lincoln, NE: U of Nebraska Press.

Hühn, Peter, et al. 2015. "*The Living Handbook of Narratology.*" Edited by Peter Hühn et al. *Theatlantic.com*. Accessed October 5. http://www.lhn.uni-hamburg.de.

Jennings, Pamela. 1996. "Narrative Structures for New Media." *Leonardo* 29 (5): 345–50.

Koenitz, Hartmut. 2014. "Five Theses for Interactive Digital Narrative." In *Interactive Storytelling: 7th International Conference on Interactive Digital Storytelling, ICIDS 2014, Singapore, Singapore, November 3-6, 2014, Proceedings*, edited by Clara Fernández-Vara, Alex Mitchell, and David Thue, 8832:134–39. Lecture Notes in Computer Science. Cham: Springer International Publishing. doi:10.1007/978-3-319-12337-0_13.

Koenitz, Hartmut. 2015. "Towards a Specific Theory of Interactive Digital Narrative." In *Interactive Digital Narrative : History, Theory, and Practice*. New York.

Koenitz, Hartmut. 2016. "Interactive Storytelling Paradigms and Representations: a Humanities-Based Perspective." In *Handbook of Digital Games and Entertainment Technologies*, 1–15. Singapore: Springer Singapore.

Koenitz, Hartmut, and Mirjam Palosaari Eladhari. 2019. "Challenges of IDN Research and Teaching." In *Technologies for Interactive Digital Storytelling and Entertainment*, 11869:26–39. 12th International Conference on Interactive Digital Storytelling, ICIDS 2019, Little Cottonwood Canyon, UT, USA, November 19–22, 2019, Proceedings. Cham: Springer International Publishing. doi:10.1007/978-3-030-33894-7_4.

Koenitz, Hartmut, Gabriele Ferri, and Tonguc Ibrahim Sezen. 2009. "Do We Need a New Narratology for Interactive Digital Storytelling? a Workshop on Theory at ICIDS 2009." In, 354.

Koenitz, Hartmut, Mads Haahr, Gabriele Ferri, and Tonguc Ibrahim Sezen. 2013. "First Steps Towards a Unified Theory for Interactive Digital Narrative." *Transactions on Edutainment X* 7775 (Chapter 2). Berlin, Heidelberg: Springer Berlin Heidelberg: 20–35. doi:10.1007/978-3-642-37919-2_2.

Laurel, Brenda. 1986. "Toward the Design of a Computer-Based Interactive Fantasy System." Ohio State University.

Montfort, Nick. 2005. *Twisty Little Passages*. MIT Press.

Murray, Janet H. 2011. *Inventing the Medium : Principles of Interaction Design as a Cultural Practice*. Cambridge, Mass: MIT Press.

Murray, Janet Horowitz. 1997. *Hamlet on the Holodeck: the Future of Narrative in Cyberspace*. New York: Free Press.

Thue, David, and Elin Carstensdottir. 2018. "Getting to the Point: Toward Resolving Ambiguity in Intelligent Narrative Technologies." In.

Wiener, Norbert. 1948. *Cybernetics or Control and Communication in the Animal and the Machine*. MIT Press.

2020a. "Stanford Encyclopedia of Philosophy." *Plato.Stanford.Edu*. Accessed June 27. https://plato.stanford.edu/.

2020b. "Wikipedia." *Wikipedia.org*. Accessed June 27. http://wikipedia.org.

# Appendix 1 Sample Entry "Transformation"

The text below is an example for a "concept" entry in the IDN encyclopedia. The structure of such an entry is

## Concept
Author name(s)

**Relation to taxonomy** (either defining a term that exist in the taxonomy or relating to closest existing term(s))

**Alternative names for concept** (if available)

**Definition**

**Explication**

**History**

**Custom sections for specific concept**

**Bibliography**

The example entry starts below

# Transformation

Christian Roth, Elisa D. Mekler

### Relation to IDN Taxonomy

Transformation is an aesthetic quality and thus an aim during the authoring process (**Authoring** > Ideation > Content > Aesthetic qualities >Transformation).

It is also an element of the experience of successful design (**Artefact** > Process > Experience > Aesthetic > Transformation)

Finally, the transformation manifests in different products (**Artefact** > Process) from the same protostory (**Artefact** > System > Protostory).

## Alternative names of concept

-

## Definition

*Transformation* is a fundamental aesthetic quality of interactive narratives with the potential to positively change interactors' thinking and behavior.

In her influential book, Hamlet on the Holodeck, Janet Murray (1997) proposes transformation as one of the characteristic pleasures of digital environments. Murray describes three distinct meanings of transformation:

- **Transformation as enactment.** The interactive narrative allows the interactor to transform themselves into someone else for the duration of the experience.
- **Transformation as variety.** The experience offers a multitude of variations on a theme. The interactor is able to exhaustively explore these kaleidoscopic variations and thus gain an understanding of the theme and different aspects.
- **Personal transformation.** The experience takes the player on a journey of personal transformation, potentially altering their attitudes, beliefs, and behaviour.

Transformation as masquerade and as variety can be seen as precursors for personal transformation (Mateas, 2001).

## Explication

"The right stories can open our hearts and change who we are" – according to Janet Murray, interactive narratives can reflect our contemporary conception of the world as multiple and enacting stories, which enables the exploration of multiple instances from different perspectives. This personal transformation is crucial for the success of commercial, artistic, and serious applications alike.

The idea of personal transformation in Interactive Digital Narrative (IDN) is based on the notion of a transformative experience, commonly defined as "causing or able to cause an important and lasting change in someone or something", in particular "causing someone's life to be different or better in some important way" (Merriam-Webster dictionary).

Murray (1997) understands transformation as a hybrid category, describing both the changes afforded by the interactor on the narrative and the changes in the interactors themselves, in their understanding of the topic at hand, thus aligning transformation closely with learning.

## History

Looking at the history of transformation and IDN, we find the term proposed by Janet Murray (1997) as one of her three aesthetic phenomenological categories for the analysis and design of interactive story experiences: immersion, agency, and transformation.

In Murray's view, a successful IDN work draws the audience in (immersion), provides agency (the satisfying ability to make meaningful changes to a virtual environment), and transforms both the virtual environment and the interactor.

In Murray's work *transformation* has distinct meanings:

**Transformation as enactment**

The IDN experience allows interactors to transform themselves into someone else for the duration of the experience, letting them experience new perspectives through roleplaying. Interactive digital narratives enable the exploration of and experimentation with different actions and consequences through the interactors' playful performances. For instance, Murray describes experiencing the same dramatic family situation – an unchanging set of events – but from multiple perspectives of each family member. Enacting – rather than merely witnessing – different characters allows experiencing such interwoven stories as one unit, which is thought to enhance interactors' capacity to imagine multiple viewpoints and explore complex processes.

**Transformation as variety**

The IDN experience offers a multitude of variations on a theme – by allowing multiple playthroughs, different playable characters, and/or affording the interactor various choices. Interactors are able to exhaustively explore these kaleidoscopic variations and thus gain an understanding of the theme.

"The kaleidoscopic power of the computer allows us to tell stories that more truly reflect our turn-of-the-century sensibility. We no longer believe in a single reality, a single integrating view of the world, or even the reliability of a single angle of perception. Yet we retain the core human desire to fix reality to one canvas, to express all of what we see in an integrated and shapely manner. The solution is the kaleidoscopic canvas that can capture the world as it looks from many perspectives – complex and perhaps ultimately unknowable but still coherent" (Murray, 1997 p. 161).

According to Murray, even a single playthrough affords a transformation by means of the choices and the subsequent awareness of paths not taken. In contrast, Michael Mateas, cocreator of the influential interactive drama Façade, restricts transformation to replay in the form of additional playthroughs (Mateas, 2001) in an attempt to reconcile Murray's perspective with Brenda Laurel's (1986) structurally strict neo-Aristotelian interpretation of ancient Greek drama.

**Personal transformation**

Personal transformation denotes the product of the game experience, potentially altering the interactor's attitudes, beliefs, and behavior. Roth and Koenitz (2016) further conceptualized personal transformation via the notion of eudaimonic appreciation, the link that connects the aesthetic presentation of an IDN to a personal dimension rooted in previous experience. Interactors construct personal meaning from engaging with a narrative. The operationalization of transformation as eudaimonic appreciation is based on entertainment theory, which differentiates between pleasure-seeking hedonic and truth-seeking eudaimonic motivations to use entertainment media (cf. Oliver & Bartsch, 2010). The exploration of aesthetic content is intrinsically motivated and understanding the meaning of an aesthetic presentation is a challenge that is driven by curiosity and identification, based on personal taste and background. Hence, aesthetic and eudaimonic appreciation is often linked to the personal meaning individuals attach to media offerings (Rowold, 2008). Empirical studies on emotionally moving and challenging experiences in games (Bopp et al., 2016; 2018) seem to largely support this notion of eudaimonic appreciation. These instances are not solely characterized by their intensely emotional impact, but players also describe these experiences as thought-provoking, personally meaningful, staying with them long after gameplay, as well as often shaping how they see and act in the world. For instance, being confronted with racism against one's character in *Skyrim* (Carver, 2011), coping with personal loss by mourning the death of one of the titular brothers in *Brothers: A Tale of Two Sons* (Fares, 2013), or dealing with ethical dilemmas in the *Mass Effect* series (Watamaniuk, Karpyshyn & Hudson, 2007).

Whitby et al. (2019) further distinguish between endo- and exotransformation. Endo- transformation refers to moments where players engage in reflection that remains limited or endogenous to the game played, resulting in changes in how the player perceives or approaches the game. Exotransformation, in contrast, denotes moments where players engage in reflection that affects their beliefs or actions outside or exogenous to gameplay.

## Custom sections

### Design Considerations

Murray called for a new set of formal conventions for handling mutability and affording transformation. While this research area is still nascent, some design considerations can be distilled from game design and learning research.

**Kaleidoscopic Storytelling**

Transformation as enactment and variety constitute opportunities for "kaleidoscopic storytelling" (Murray, 1997), the creation of coherent multisequential – often contrasting – narratives, which require the interactor to construct their own meaning and acknowledge the multiplicity of perspectives. *Naughty Dog's The Last of Us 2* (Druckmann, Newman & Margenau, 2020), for example, has players enact characters from two different warring factions, thereby providing players an opportunity to engage with the trauma and pain of both. *PeaceMaker* (Sweeney, Brown & Burak, 2007) similarly allows players to assume the role of the Palestinian president or Israeli prime minister, to juxtapose opposing perspectives and reflect the political complexities.

**Sequence Questions**

Drawing from film theory, Dena (2017) introduces "sequence questions" to guide interactors' experience: Sequencing divides the intended experience into a series of questions for interactors to consider. An overall question is introduced at the beginning and perhaps answered near the end. However, multiple short-term questions keep driving the interactors' interest, and offer them the opportunity to imagine many possible outcomes before the narrative resolution. Sequence questions are formulated from the perspective of the interactor, but grounded in the actions of the player character (e.g., in *Papers, Please* (Pope 2013): "What will happen to my family?", "Which faction can I trust (and can I trust anyone)?", "Can I let this person enter the country? What are the repercussions of my actions?").

**Disorienting dilemma**

Within his Transformative Learning Theory, Mezirow (1991) introduces the "disorienting dilemma" which he associates with major life crises or transitions. However, interactive narrative experiences designed to promote transformation, can also challenge interactors to change their frames of reference by re-evaluating their assumptions and beliefs and, perhaps, making a conscious decision to redefine their world view. Dilemmas can also be rooted in moral intuitions, such as having individuals encounter moral violations in IDN in which players must decide quickly how to respond.

**Example: Interactive Installation "Angstfabriek"**

The Angstfabriek (Dutch for fear factory) is an educational installation in the form of a complete building and lets visitors experience fear-mongering and the related safety industry, with the goal of eliciting reflection, insight and discussion.

Interactors visit the installation without a predefined role, but switch into the new role of whistleblowers during the experience. This 'transformation as masquerade' is designed to allow for a perspective change: to see behind the curtain of fear-mongering and the industry benefiting from it.

Roth (2019) sees transformation as a deeply personal and meaningful insight with the potential to change attitudes and behavior and evaluated the potentially transformative user experience of this installation via a focus group interview and a pilot user experience study (N = 32). The research revealed the importance of sufficient scripting of visitors regarding their role and agency, highlighting the conceptual connection between interactive digital narrative design and interactive theater design. Interactors did not know what to expect, what role to inhabit and how to perform it within the unknown rules and constraints of the interactive installation.

Working undercover in the factory, visitors could either comply with the director's assignments or decide to actively sabotage the production of fear-inducing media messages. Only a few visitors made use of the disruptive roleplaying and barely sabotaged. Thus, the role of whistleblower remained underdeveloped and mainly determined by the visitors' imagination. Similarly, reflection on the experience was not fostered by design and it was up to the visitors to meet and discuss their experience afterwards.

## Evaluation Tools

Transformation is often equated with behavioral and/or attitudinal change, which is reflected in evaluation metrics. However, these measures do not clearly specify how this change came about or whether it will endure in the long-term. Relative to other dimensions, transformation as an aspect of the user experience has received relatively little attention to date.

Roth has captured learning and transformation implicitly as an aspect of his granular framework to evaluate the user experience of interactive digital narratives (Roth, 2016). The framework has since been explicitly mapped to Murray's category of transformation (Roth & Koenitz, 2016). Specifically, Roth's measurement toolkit captures the dimension of transformation via eudaimonic appreciation.

Roth's measurement of eudaimonic appreciation uses a 5-item scale adapting and extending the work of Rowold (2008) and Cupchik and Laszlo (1994). After experiencing an artefact, interactors rate statements on a 5-point Likert scale, ranging from "strongly disagree" to

"strongly agree". Typical items are "The experience made me think about my personal situation.", "The experience told me something about life." and "The experience moved me like a piece of art.". The resulting values are aggregated and form a 'transformation score' with 1 being the lowest and 5 being the highest. While these values differ between individuals - based on their experience, expectations, and preferences - the resulting mean score gives an indication of the overall rating of an artefact in regard to its transformation quality. As an addition to this quantitative measurement, interactors are asked within a series of qualitative questions to elaborate on their experience and the meaning it had to them. This data allows for a content analysis revealing underlying patterns of a meaningful experience.

Understanding the different levels of meaning-making is crucial for the design of transformative experiences. While players interact with an IDN system, they are continuously extracting information to understand past and present events and to plan their actions. The interpretation occurs in terms of narrative game mechanics and their meaning in relation to the narrative theme. Roth,

Van Nuenen, and Koenitz (2018) proposed a model to conceptualize the narrative meaning-making processes in the form of a double-hermeneutic circle (cf. Gideons 1987; Veli-Matti Karhulahti, 2012): a) when interacting with the system and b) when interpreting the instantiated narrative at any point of the experience. This model allows the analysis of IDN works as a shared meaning-making activity between designer and audience.

Other works (Mekler et al., 2018) have applied Fleck & Fitzpatrick's reflection framework to evaluate transformative reflection from qualitative interview data. Via this approach it is possible to (1) identify evidence of reflective thought in interactors' accounts, (2) distinguish it more clearly from non-reflective forms of thinking, as well as (3) differentiate between levels of reflection. However, the effectiveness of this approach is determined by how well interview participants are able to articulate their experience.

**Related concepts: Elements impacting transformation**

(links to other encyclopedia entries):

- Usability
- Immersion
- Agency
- Embodiment
- Identification
- Self-Determination Theory (Competence, Autonomy, Social Relatedness)
- Motivation
- Media and Transmedia literacy
- Technological barriers and opportunities

## Connected theories

The **Transformative Learning Theory** developed by Jack Mezirow sees transformative learning as the critical awareness of unconscious suppositions or expectations and the evaluation of their relevance for making an interpretation.

Mezirow's theory can be connected to Murray's understanding of personal transformation and gives important insight on how such transformations can form.

Mezirow suggests that we interpret meaning out of experiences through "a lens of deeply held assumptions" (Mezirow, 1991). When people are confronted with information or experiences that contradict or in some way challenge their understanding of the world, they are prompted to reassess their assumptions and processes of meaning-making. This can lead to a fundamental change of the learner's world view, shifting from an unconditional acceptance of available information into a conscious and reflective way of learning that supports real change. Mezirow refers to potentially transformative experiences as "disorienting dilemmas", which includes events such as loss, trauma, stressful life transitions or other interruptions.

Mezirow states three dimensions for the process of "perspective transformation": psychological (changes in understanding of the self), convictional (revision of belief systems), and behavioral (changes in lifestyle). This type of learning transcends simply acquiring knowledge as it concerns deep, useful and constructive learning which offers constructive and critical ways for learners to consciously give meaning to their lives.

**Reflection Framework** by Fleck and Fitzpatrick (2010) outlined a framework of different stages of reflection. While drawing from Mezirow and intended for Human-Computer Interaction, there is considerable overlap between their individual reflection stages and Murray's notions of transformation: I.e., Descriptive reflection denotes reasoning and/or justification, but without exploration of alternate explanations or perspectives. Dialogic reflection, in contrast, entails looking for relationships between instances of experience, considering alternate explanations and perspectives, as well as questioning, hypothesizing and interpreting about one's experiences. As such, it bears much semblance to the notions of transformation as enactment and variety.

Transformative reflection, in turn, refers to moments when the reflector's original point of view is somehow altered or transformed to take into account the new perspectives they just explored.

Finally, critical reflection denotes relating one's experiences to wider socio-cultural and ethical implications. Hence, both transformative and critical reflection constitute examples of personal transformation.

# Bibliography

Bopp, J. A., Mekler, E. D. & Opwis, K. (2016). Negative Emotion, Positive Experience: Emotionally Moving Moments in Digital Games. CHI 2016. https://doi.org/10.1145/2858036.2858227

Bopp, J. A., Opwis, K. & Mekler, E. D. (2018). "An Odd Kind of Pleasure" Differentiating Emotional Challenge in Digital Games. *CHI 2018*. https://doi.org/10.1145/3173574.3173615

Laurel, B. (1986). Toward the Design of a Computer-Based Interactive Fantasy System. Ohio State University.

Carver, G., (2011). *Skyrim*. Bethesda Game Studios.

Chen, J. C., & Martin, A. R. (2014). Role-Play Simulations as a Transformative Methodology in Environmental Education. Journal of Transformative Education, 13(1), 85–102. doi:10.1177/1541344614560196

Cupchik, G. C. and Laszlo, J. (1994). The landscape of time in literary reception: Character experience and narrative action. Cognition & Emotion, 8(4):297–312.

Dena, Christy. (2017). "Finding a Way: Techniques to Avoid Schema Tension in Narrative Design" ToDiGRA. doi:10.26503/todigra.v3i1.63

Druckmann, N., Newman, A. and Margenau, K., (2020). The Last of Us 2. Naughty Dog.

Fares, J., (2013). Brothers: A Tale Of Two Sons. Starbreeze Studios.

Fleck, R. & Fitzpatrick, G. (2010). Reflecting on Reflection: Framing a Design Landscape. OzCHI 2010. http://dx.doi.org/10.1145/1952222.1952269

Giddens, A. (1987).Social Theory and Modern Sociology. Stanford University Press.

Karhulahti, V.-M. (2012). Double fine adventure and the double hermeneutic videogame (pp.19–26). Presented at the Foundations of Digital Games, New York, New York, USA: ACM Press. http://doi.org/10.1145/2367616.2367619

Koenitz H., Di Pastena A., Jansen D., de Lint B., Moss A. (2018) The Myth of 'Universal' Narrative Models. In: Rouse R., Koenitz H., Haahr M. (eds) Interactive Storytelling.

ICIDS 2018. Lecture Notes in Computer Science, vol 11318. Springer, Cham. https://doi.org/10.1007/978-3-030-04028-4_8

Mateas, M. (2001). A Preliminary Poetics for Interactive Drama and Games. Digital Creativity, 12(3), 140–152. http://doi.org/10.1076/digc.12.3.140.3224

Mekler, E. D., Iacovides, I. & Bopp, J. A. (2018). "A Game that Makes You Question" Exploring the Role of Reflection for the Player Experience. *CHI PLAY 2018*. https://doi.org/10.1145/3242671.3242691

Mezirow, J. (1991): Transformative Dimensions of Adult Learning. The Jossey-Bass higher and adult education series. Jossey-Bass, San Francisco

Merriam-Webster. (n.d.). Transformative. In Merriam-Webster.com dictionary. Retrieved August 4, 2020, from https://www.merriam-webster.com/dictionary/transformative

Murray, J.H. (1997, 2016): Hamlet on the Holodeck: The Future of Narrative in Cyberspace. Free Press, New York

Oliver, M. B. & Bartsch, A. (2010). Appreciation as audience response: Exploring entertainment gratifications beyond hedonism. *Human Communication Research* 36, 1, 53–81. http: //dx.doi.org/10.1111/j.1468-2958.2009.01368.x

Pope, L. (2013). Papers, Please. 3909 LLC.

Roth, C. (2016). Experiencing Interactive Storytelling. Vrije Universiteit Amsterdam. https://research.vu.nl/en/publications/experiencing-interactive-storytelling

Roth, C. (2019). The 'Angstfabriek' Experience: Factoring Fear into Transformative Interactive Narrative Design. In International Conference on Interactive Digital Storytelling (pp. 101-114). Springer, Cham.

Roth, C. & Koenitz, H. (2016). Evaluating the user experience of interactive digital narrative. In Proceedings of the 1st International Workshop on Multimedia Alternate Realities. 3136.

Roth, C., Van Nuenen, T. & Koenitz, H. (2018). Ludonarrative Hermeneutics: A Way Out and the Narrative Paradox. In R. Rouse, H. Koenitz and M. Haahr (eds.). Interactive Storytelling - 11th International Conference on Interactive Digital Storytelling, ICIDS 2018, Dublin, Ireland, December 5-8, 2018, Proceedings. Cham: Springer. 93–106.

Rowold, J. (2008). Instrument Development for Esthetic Perception Assessment. Journal of Media Psychology, 20(6):35–40.

Sweeney, T., Brown, E. and Burak, A., (2007). Peacemaker. ImpactGames.

Watamaniuk, P., Karpyshyn, D. & Hudson, C., (2007). Mass Effect. BioWare.

Whitby, M. A, Deterding, C. S. & Iacovides, I. (2019): "One of the Baddies All Along" Moments that Challenge a Player's Perspective. CHI PLAY 2019. https://doi.org/10.1145/3311350.3347192